# 10 Questions to Fall in Love with ChatGPT: An Experimental Study on Interpersonal Closeness with Large Language Models (LLMs)

**Jessica Szczuka, PhD\***
Jessica.Szczuka@uni-due.de
INTITEC Research Group/ Research Center Trustworthy Data Science and Security
Bismarckstr. 120 - 3.1.OG.3320
University Duisburg-Essen
47057 Duisburg, Germany

**Lisa Mühl, M.Sc.**
lisa.muehl@uni-due.de
INTITEC Research Group
University Duisburg-Essen
Duisburg, Germany

**Paula Ebner, M.Sc.**
paula.ebner@uni-due.de
INTITEC Research Group
University Duisburg-Essen
Duisburg, Germany

**Simon Dubé, PhD**
simondube.ta@gmail.com
Kinsey Institute
Indiana University
Bloomington
USA

\* Corresponding Author

**Dr. Jessica Szczuka** is the Head of INTITEC (Intimacy with and Through Technologies) and a leading expert in digitized intimacy. She earned her Ph.D. in Social Psychology: Media and Communication from the University of Duisburg-Essen. Specializing in human-robot interaction (HRI) and human-computer interaction (HCI), her research explores how people form social connections with AI systems, particularly in intimacy and sexuality.

**Lisa Mühl,** M.Sc., is a Ph.D. candidate at INTITEC research group, which she joined in October 2022. She holds a master's degree in Applied Cognitive and Media Science, a combination of computer science and psychology. Her doctoral research is conducted as part of SENTIMENT, an interdisciplinary project funded by the German Federal Ministry of Education and Research, focusing on intimate communication and privacy concerns in natural language dialogue systems.

**Paula Ebner**, M.Sc., is a Ph.D. candidate at INTITEC. She earned her bachelor's degree in psychology from Maastricht University, where she began her work in sex research, focusing on sexual synchrony and functioning. After completing a research master's in legal psychology at the same institution, she joined INTITEC in 2023. For her dissertation, she is investigating how romantic and sexual fantasies influence parasocial relationships with chatbots.

**Simon Dubé, PhD**, is a Research Fellow at the Kinsey Institute (Indiana University, Bloomington) and Affiliate Professor in the Department of Sexology at the Université du Québec à Montréal. His research focuses on erobotics – the study of human-machine erotic interaction and co-evolution. His work also explores space sexology and the integration of sex research into space programs.

**Notes.**

Dr. Simon Dubé consults and acts as Scientific Director for the International Sexual Health and Wellness Research Institute (le Shaw; Byborg Enterprises).

The recruitment was funded by the Research Center Trustworthy Data Science and Security (University Alliance Ruhr, Germany, https://rc-trust.ai/).

**Abstract**

Large language models (LLMs), like ChatGPT, are capable of computing affectionately nuanced text that therefore can shape online interactions, including dating. This study explores how individuals experience closeness and romantic interest in dating profiles, depending on whether they believe the profiles are human- or AI-generated. In a matchmaking scenario, 307 participants rated 10 responses to the Interpersonal Closeness Generating Task, unaware that all were LLM-generated. Surprisingly, perceived source (human or AI) had no significant impact on closeness or romantic interest. Instead, perceived quality and human-likeness of responses shaped reactions. The results challenge current theoretical frameworks for human-machine communication and raise critical questions about the importance of authenticity in affective online communication.

**10 Questions to Fall in Love with ChatGPT: An Experimental Study on Interpersonal Closeness with Large Language Models (LLMs)**

Recent progress in Large Language Models (LLMs) are enabling the advent of systems with natural language capabilities that rival those of humans. Communication being at the foundation of our social relationships, these technological advancements are reshaping both how and with whom humans can establish intimate connections (Dubé & Anctil, 2020). These systems can serve several functions, such as practicing sociosexual interactions (e.g., flirting), enhancing communications (e.g., crafting more compelling texts), and powering artificial companions (cf., Nomi.AI, 2024; Replika, 2024). First surveys found that some people already want to use AI to craft love letters and struggle to distinguish between those written by AIs and humans (McAfee, 2023). Moreover, the $5^{th}$ most downloaded smartphone application in 2024 was RIZZ, which is an LLM-powered so-called "dating-assistant" that is built to "generate personalized responses that are sure to impress your crush" (Miller, 2024; Predin, 2024).

What may seem like an easy path to success in online dating raises questions about the importance of authentic and unique communication, particularly in a context where genuine self-presentation is key, rather than relying on messages that merely reflect statistically optimized responses. This issue is especially pressing as meeting partners online is one of the most prevalent ways to get to know new people and find love. Given the potential to disrupt the importance of authenticity in online dating communication, this study specifically investigates whether the source of a message, human or artificial, affects the development of positive first impressions, measured through interpersonal closeness and intimate interest.

To do so, this study employed a mock AI-matchmaking scenario based on an adapted version of the Interpersonal Closeness Generation Task (ICGT, commonly referred to as the *36 questions to fall in love,* Aron et al., 1992). In this version, participants first provided their own answers to 10 of the ICGT questions. Following this, they were presented with answers to the same set of questions by their allegedly "perfect match" (i.e., a human or AI partner, similarly to a dating profile), mimicking a two-way interaction, similarly to one you may find on dating or matchmaking websites. Importantly, regardless of whether they were presented as coming from a human or AI, all the answers from this "perfect match" were generated by the freely available text-generative chatbot, ChatGPT 3.5. and did not differ between the two conditions (OpenAI, 2023). Figure 2 shows an exemplary answer. In doing so, this study simultaneously contributes to a deeper understanding of whether people could develop feelings of interpersonal closeness and intimate interest toward LLM-generated text if it was indistinguishable in text-based communications (i.e., via the human partner condition, which is LLM-generated), as well as whether the awareness of the human or artificial origin of a personal description influences the capacity to evoke such feelings towards the described entity.

**Literature Review**

*Social Reactions to Artificial Entities: Potential Theoretical Frameworks*

Human communication and its related behaviors are shaped by interpretations of the communication partner and heavily influenced by the nature of participants (Pavitt & Proud, 2009). While interhuman interactions involve partners belonging to the same ontological category, human-machine interactions cross categories (Dautenhahn, 2004). Traditional ontological boundaries, such as origin, autonomy, intelligence, and emotion have historically distinguished humans from machines (Guzman, 2020). Until recently, communication was thus shaped by an anthropocentric expectancy bias, where partners were assumed to be human or humanlike. Encounters with less humanlike entities often created a sense of disruption, as individuals instinctively applied human communication patterns to machines (Edwards et al., 2019). However, with the rapid advancements in AI (particularly LLMs) these boundaries are increasingly blurred, challenging established user perceptions and communication paradigms. For example, Guzman (2020) demonstrated that shifting perceptions of human-machine differences shape decisions to engage with technology, preferences for human versus machine interactions, and expectations for communicative roles. The study illustrated that people perceive multiple boundaries between humans and AI, with emotion being a key distinction, though interactions with emotion-emulating AI, have led some to reconsider this divide. Intelligence is also debated, with differing views on whether humans or computers are "smarter", as some see human nature as inherently flawed compared to AI. As technology increasingly mimics human traits, these boundaries evolve. The blurring of ontological boundaries further challenges traditional theories of human-machine interaction, such as CASA (Nass & Moon, 2000) and the Media Equation Theory (Reeves & Nass, 1996), which rely on clear distinctions between human and non-human agents.

Media Equation Theory proposes that humans tend to react socially to technology if it is equipped with social cues (Nass & Moon, 2000; Reeves & Nass, 1996). In this paradigm, computers are regarded as social actors, leading users to interact with them as if they were human, despite being fully aware that these machines lack "selves" or human motivation (Nass et al., 1993; van der Goot & Etzrod, 2023). An extensive body of both theoretical and empirical research has demonstrated that social responses towards non-human, artificial systems are comparable to those directed towards other humans (cf., Hoffmann et al., 2009; Horstmann et al., 2018; Johnson et al., 2004). People tend to apply human social scripts to computers automatically, even with minimal cues like simple text outputs (Nass & Moon, 2000; Reeves & Nass, 1996). This can lead users to rely on human decision rules, even when they know these are not appropriate for assessing machine behavior (Nass et al., 1993; van der Goot & Etzrod, 2023). Media Equation Theory further suggests that when technology displays social cues, such as interactivity, natural language, and the fulfillment of a social role, the perception of them as social actors is enhanced (Nass & Moon, 2000; Reeves & Nass, 1996). These components are deliberately integrated into new LLM applications, enabling conversational agents to better understand human language and prompting users to perceive them as genuine actors (Sarigul et al., 2024). Although earlier research (Fox & Gambino, 2021) argues that the limited sophistication of human-like cues in artificial agents made interpersonal relationship models largely inapplicable, evolving machine-like heuristics now increasingly allow conversational agents to act as more complex social actors. This shift enables users to assign roles to these agents, fostering a sense of closeness and mimicking or allowing interpersonal and intimate relationships (Sarigul et al., 2024). In a time when technology is increasingly indistinguishable from human beings, it thus becomes necessary to reconsider the suitability of the Media Equation Theory (van der Goot & Etzrod, 2023). While research has demonstrated that technologies can elicit various social reactions, supporting the theory's application, most empirical studies have focused on scenarios where users, upon reflection, are aware that they are interacting with non-human entities (e.g., because they were interacting with an embodied robot which is clearly a non-human entity; cf., Hoffmann et al., 2009; Horstmann et al., 2018). Yet, it remains uncertain how users would respond when they are unable to differentiate between humans and machines or the content each generates.

In this context, the Media Evocation paradigm, introduced by van der Goot and Etzrod (2023), offers a complementary perspective. It categorizes machines as entities existing in a liminal space between traditionally opposing concepts, such as person and object. Within this paradigm, computers are framed as "evocative objects" (Turkle, 1984, 2005) that provoke self-reflection, challenge categorization, and blur the boundaries of mind and identity. Positioned "betwixt and between", these machines prompt users to question the nature of social actors, demonstrating that such actors need not be human. This approach shifts the focus from instinctive responses to a deliberate process of reflection, where users negotiate the nature of machines, themselves, and their relationship. By embracing the paradoxical nature of machines, the Media Evocation paradigm inspires users to rethink established assumptions and frameworks (van der Goot & Etzrod, 2023).

As a complement to the Media Evocative paradigm, willing suspension of disbelief and the Sexual Interaction Illusion Model (SIIM) can respectively provide valuable perspectives on how and why knowing that a text was generated by a non-human could still be of interest for humans. Both approaches do not argue that social reactions towards artificial entities are mindless but rather a product of active engagement. More specifically, willing suspension of disbelief proposes that we may choose to enjoy something despite or perhaps even because of knowing its fictitious nature (originally postulated in the context of theater perceptions, Aylett et al., 2023; Coleridge, 1984), whereas the SIIM suggests that some people may engage specifically in sexualized interactions with artificial entities for hedonistic reasons. Both frameworks have in common that individuals accept and/or embrace the interactions as they are, enjoying the experience without dwelling on the artificial nature of the interaction or persona (Szczuka et al., 2018). While these theoretical frameworks provide valuable context for understanding why individuals engage with artificial entities once they are aware of their origin, the present study takes place in a dating context where users may not actively seek interaction with artificial agents. This makes the Media Equation Theory and the previously discussed constructs a more suitable theoretical foundation, while still allowing for insights that could further elaborate on the study's potential findings.

*Empirical Insights*

Research on interpersonal closeness and intimate interests toward humans compared to artificial entities (e.g., chatbots, virtual companions, robots) is scarce, but provides additional insights into why people may feel drawn to artificial partners or not. It suggests that, should one know the nature of an agent, they prefer humans over artificial partners—especially, in romantic and/or sexual contexts (Dubé et al., 2022; Szczuka, 2022; Szczuka & Szymczyk, 2024). For instance, an empirical study comparing the flirtatiousness of voice assistants to identical content delivered as a voice message from a human showed that the human voice was perceived as more sexually attractive and flirtatious (Szczuka, 2022). The author argued that flirtatiousness signals a human drive and interest towards others that can only be (perceived as) authentically felt by another human. Humans may also prefer other humans for reproductive purposes, societal norms discouraging non-human intimacy, as well as the current shortcomings of technologies that prevent more complex erotic interactions (e.g. Banks & van Ouytsel, 2020; Szczuka et al., 2018).

Human relationships with artificial personas, including technologies, such as virtual companions, chatbots, or robots, are often portrayed and discussed as a "last resort" for individuals who struggle with social interactions with other humans, including but not limited to those with intellectual and physical disabilities or neurodiverse individuals (Döring & Poeschl, 2019; Turkle, 2011). Rooted in third-person effect, this may trigger responses based on social desirability, and the reflex of separating oneself from such stereotype, manifesting in negative reactions towards intimate technologies (Mühl & Szczuka, 2024; accepted). First evidence of such stigma was empirically demonstrated by Dubé and colleagues (2023) who found that, the more human-like the erotic technology (including virtual partners), the more perceived stigma is elicited by its use.

On the contrary, there is also research which addresses people who seek artificial partners precisely for their advantages and the unique relationships that they may provide, fully aware of their companion's nature (Lievesley et al., 2023). Pentina and colleagues (2023) recently conducted one of the first multi-method studies with users of the conversational agent Replika and found that human-likeness and authenticity, meaning, perceived agency, autonomy, and uniqueness, play a key role in accepting an AI-driven interaction partner. Additionally, artificial agents can facilitate continuous access to judgment-free communication and intimacy (Dubé & Anctil, 2020). This can include communication about various romantic and sexual desires, which may deviate from traditional or conservative sociosexual scripts and norms, and could otherwise be kept secret to avoid societal backlash (Dubé & Anctil, 2020).

However, it must be noted that the majority of the previously mentioned studies examined scenarios in which users were fully aware of the ontological nature of their communication partner (e.g., because they were embodied robots or because of repeated interactions with chatbots). The emergence of large language models (LLMs), on the other hand, enables interactions in which individuals may not be aware of whether they are engaging with another human or an AI-generating text. This shift raises important questions about how dating-related texts are perceived when generated by LLMs versus when they originate from a human, emphasizing the need to explore whether and how users distinguish between human- and AI-produced communication.

*Influencing Variables*

To gain a more comprehensive understanding of how individuals respond to dating profiles created by either a human or an AI, this study also incorporates key potential influencing variables. A review of the literature suggests that gender and cultural values and dimensions may play a role in shaping these reactions. The inclusion of cultural factors was particularly informed by the researchers' observations of sample differences across Germany and Canada, highlighting the relevance of cross-cultural perspectives in this context.

In terms of gender, one of the most robust findings stemming from the burgeoning body of research on sexualized interaction technology is that men are more interested in intimacy with machines than women (e.g., sex (with) robots Bame, 2017; Brandon & Planke, 2021; Brandon et al., 2022; Dubé et al., 2023; Dubé et al., 2024; Hanson, 2013; Nguyen, 2020; Nordmo et al., 2020; Oleksy & Wnuk, 2021; Scheutz & Arnold, 2016). This may stem from a combination of gynoid marketing targeting heterosexual men, media representations mostly depicting men-gynoid intimate relationships, societies devaluing women's sexuality while valuing men's, as well as discrepancies in sexual drive, risk-taking, and sensation seeking (cf., Dubé et al., 2022; Peixoto, 2023). Regardless of the reason, this may translate into a propensity for men to experience greater feelings of interpersonal closeness and interest in intimate interactions with both human and artificial partners than women in this study. In parallel, differences grounded in location, culture, and cultural values/dimensions, such as those that may exist between North American and European countries, like Germany and Canada, may further influence people's reaction toward human-machine intimacy. In that regard, and acknowledging that both are multicultural societies, Hofstede's cultural dimensions (Hofstede & Bond, 1984) indicate that while Germany and Canada exhibit some similarities, they present differences in indulgence-restraint (i.e., the extent to which they attempt to control their desires and impulses) and uncertainty avoidance (i.e., their discomfort with ambiguous, unknown, or unfamiliar situations). These differences may influence Germans´ and Canadians´ respective reactions to artificial partners, particularly in matters of intimacy and sexuality. In practice, this may manifest by Canadians being generally more accepting of new ideas or products and more willing to try something new, while Germans tend to enjoy detailed and well-thought-out concepts. Applied to the context of potential intimate communication with artificial partners, these differences between Germans and Canadians on Hofstede's cultural dimensions would likely translate into Canadians being more open to forming interpersonal closeness and intimate interest in machines than Germans.

While the variables mentioned above were included to examine potential interaction effects, the present study also aimed to incorporate additional influential variables as covariates in the main analyses. Based on the literature review, these variables included loneliness, attitudes toward artificial intelligence, liking of the response content, and the participant's age. However, not all of these variables will be included in the final analyses due to methodological and statistical considerations, such as multicollinearity concerns, the need to maintain a balanced model complexity, and the empirical relevance of each variable in this context. Additionally, preliminary analyses may indicate that some variables do not significantly contribute to explaining variance in the dependent measures, making their inclusion unnecessary for the main calculations (see beginning of the methods section).

**The Current Study**

Leveraging tools available at the start of this project, this study primarily aimed to experimentally test whether people could develop interpersonal closeness and intimate interest toward a potential partner based in information generated by a LLM (OpenAI, 2023), which they believe either came from a human or an AI partner (**objective 1**). Additionally, the study aimed to explore this phenomenon across genders and cultural values/dimensions (i.e., German and Canadians; (**objective 2**). Based on the previous research described, it was hypothesized that participants would report greater closeness and interest in intimate interactions when they believe that a dating profile is representing another human compared to an AI partner (**hypothesis 1**). It was also hypothesized that, compared to women, men would report greater interpersonal closeness and interest in intimate interactions for both the profiles that are perceived to be written by another human, as well as those written by an AI partner (**hypothesis 2**). Lastly, it was hypothesized that Canadians would report greater closeness and interest in intimate interactions for the AI partner, with no difference for the human partner, compared to Germans (**hypothesis 3**). Additional exploratory analyses included loneliness (Hays & DiMatteo, 1987), attitudes toward artificial intelligence (Schepman & Rodway, 2023), liking of the response content, and the participant's age to gain deeper insights into the underlying mechanisms that influence how individuals develop a sense of closeness to messages, whether generated by an AI.

## Method

This study was approved by ethics Committee of the Faculty of Computer Science, University Duisburg-Essen, Duisburg, Germany (ethics certification number: 2304SPSJ5267). Informed consent was obtained electronically from each participant prior to study completion. Preregistration, data, and material are available on the Open Science Framework (https://osf.io/8gefb).

### Participants

A power analysis was conducted using G*Power to determine the required sample size. The objective was to achieve a power of .95 to detect an effect size of $f2(V) = .04$ (small to medium effect) with an alpha level of $α = .05$. The recommended sample size of the utilized two-way multivariate analysis of variance and covariance (MANCOVA) was found to be 347. A convenience sample of 274 Canadians and 269 Germans was initially recruited on Prolific in June 2023. Multiple attention and manipulation checks were performed to ensure participants paid attention during the study. The answers of 101 German and 91 Canadian participants had to be removed from the analysis because they failed the manipulation check, incorrectly identifying the ontological class of their match. Because the perceived source of the profile is the key to the study, this hard cut-off needed to be made to ensure that people were aware of the assigned condition. In total, 142 German (46.2%) and 165 Canadian participants (53.8%) were included in the analysis. The final sample contained 163 males and 144 females aged 19 to 79 ($M = 34.03$, $SD = 10.76$).

**Procedure**

To test the hypotheses, a 2 (vignette: human/artificial persona) x 2 (participant gender: male/female) x 2 (participant cultural values/dimensions: German/Canadian) between-subject study was designed. This section first outlines the experimental design, followed by an explanation of how the Interpersonal Closeness Generating Task (ICGT) (Aron et al., 1997) was implemented in this context.

This study framed as an "online dating/matchmaking scenario", in which participants were told they would engage with a novel proprietary matchmaking algorithm, designed to pair them with a compatible potential partner. The central task revolved around 10 self-disclosure questions adapted from the ICGT.

Participants were informed that the study aimed to test a novel proprietary matchmaking algorithm, which is designed to find and match them with potential intimate partners. They were requested to provide information about their age, personality, and preferred match (i.e., gender, gender identity, and sexuality of people they are interested in). They were then instructed to write their own answers to the 10 ICGT questions that were listed, simulating the self-disclosure process.

Once they were done, participants were informed that they had been matched with either an AI or a human partner (between design). They were then shown the ICGT responses of their assigned match (human or AI persona) and asked to rate the appeal of each response.

Critically, participants did not engage in a live interaction with their match. Instead, they were randomly assigned to one of two conditions and explicitly informed that they had been matched with either an AI partner or a human partner. Regardless of condition, they were then presented with prewritten responses attributed to their assigned match. For clarification, they were not told that the partner is "live" but rather that this is an existing profile that was matched, comparable to other dating applications.

Subsequently, participants were asked to answer questions regarding their level of interpersonal closeness and interest in engaging in further interactions with the matched partner. Following this, participants were asked to answer additional measures (e.g., loneliness) and demographics (see next section for details). At the end of the study, participants received a debriefing and their compensation.

Figure 1 provides an overview of the experimental design, while Figure 2 presents an example of how the responses from the human or non-human partner were displayed.

**Measures**

*Inclusion of the Other in the Self*

To assess the closeness between the participant and the chosen match, the Inclusion of Other in the Self (IOS) Scale (Aron et al., 1992), a single-item measure, was used. Participants were shown seven pairs of circles, varying from barely touching to almost fully overlapping. One circle was labeled "me" and the other one was labeled "partner." Participants were asked to choose the pair that best represented their perceived connection with the match.

*Interest in Intimate Interactions*

To assess the interest in intimate interactions with the assigned match, a four-item measure from Dubé et al. (2022) was employed, which measures the interest in love, friendship, sexual encounter, and intimate interaction. The items were asked to be rated on a 5-point Likert scale (1 = not at all, 5 = definitely) and were found to have good internal consistency ($\alpha = .86$).

*Loneliness*

To assess the feeling of loneliness among participants, the short form of the UCLA Loneliness Scale (ULS-8; Hays & DiMatteo, 1987) was used. The scale contains eight items (e.g., "I lack companionship" or "I feel left out"). The study employed a 5-point Likert scale with values ranging from 1 = never to 5 = always. As no validated German version of the ULS-8 is available, the items were translated via cross translation. The internal consistency of the measure was $\alpha = .90$.

*Attitude towards Artificial Intelligence*

Attitude towards AI was measured using the General Attitudes towards Artificial Intelligence Scale (GAAIS; Schepman & Rodway, 2023), which measures positive (12 items, e.g., "Artificial Intelligence is exiting") and negative (8 items, e.g., "I think Artificial Intelligence is dangerous) attitudes towards AI. The study employed a 5-point Likert scale with measures ranging from 1 = strongly disagree to 5 = strongly agree. The German version of the GAAIS was retrieved from Sindermann et al. (2021). The internal consistency of the items was $\alpha = .90$.

*Anthropomorphism*

Anthropomorphism, the attribution of human qualities to nonhuman entities, was assessed using the Godspeed questionnaire (Bartneck et al., 2009). This semantic differential scale asks participants to rate profiles as machinelike or humanlike, and artificial or lifelike. The German version of the questionnaire was obtained from (Bartneck, 2018).

*Demographics*

Age was assessed with an open-ended question, while gender and sex were assessed using both open and multiple-choice items.

*Note that additional measures were used, but not included in the following analysis (e.g., mate value and Mini-IPIP).

**Manipulation Check of the Stimulus**

A manipulation was conducted to ensure that the vignette was successful and that the conditions were perceived to be different from each other. The manipulation check showed significant differences between the AI and human conditions on the four anthropomorphism items of Bartneck's Godspeed questionnaire (Bartneck et al., 2009). People who were matched with an AI persona rated the answers as less natural ($t(315) = -3.182\ p < .001$), human-like ($t(315) = -3.375\ p = .004$), conscious ($t(315) = -4.673\ p < .001$), and lifelike ($t(315) = -2.863\ p < .001$) than those who were supposedly matched with a human. Those participants rated the answers as more natural, human-like, conscious, and more lifelike. Table 1 shows all relevant descriptives.

## Results

**Effects of Ontological Class, Gender, and Cultural Values/Dimensions on Inclusion of the Other in the Self and Intimate Interest**

The following chapter presents the main calculations using a two-step approach. First, relevant covariates are identified, followed by the main-effect analysis (H1), which includes an examination of interaction effects (H2 & H3).

To decide the covariate for the following analyses instead of overloading the calculation, two multiple linear regression analyses were conducted to determine which of the additional measures of loneliness, attitudes toward artificial intelligence, liking the content of the answers, and the participant's age, predicted closeness and interest in intimate interactions best. The model predicting closeness to the match came out significant, $R^2 = .518$, $F(4, 302) = 80.981$, $p < .001$ (see Table 2), as well as the model predicting interest in intimate interactions, $R^2 = .467$, $F(4, 302) = 66.234$, $p < .001$ (see Table 3). The sole significant predictor in both models was how much participants liked the content of the match's answers ($R^2 = .467$, $F(4, 302) = 66.234$, $p < .001$ for closeness to the match and $p < .001$ for interest in interpersonal interactions). Consequently, this measure of "liking the answers" served as the covariate.

A two-way multivariate analysis of variance and covariance (MANCOVA) was performed to investigate if the gender and cultural values/dimensions of the participants, as well as the ontological class of the match affected the ratings of closeness and interest in intimate interactions with their match. The assumption of covariance homogeneity was met with a Box´s M value of 20.50 and a corresponding *p-value* of *.519.* Similarly, the assumption of variance homogeneity was met, as Levene's Test for Equality of Variances was significant for both dependent variables ($p = .816$ for closeness, and $p = .132$ for interest in intimate interactions).

There was no significant main effect of the ontological class of the match on the combined dependent variables [$F(2, 297) = .070$, $p = .933$]. Although the descriptive data revealed that participants matched with a human partner gave slightly more positive ratings on both dependent measures compared to those matched with an AI persona, this difference did neither reach significance for either the closeness measure [$F(1, 298) = .114$, $p = .736$], nor for the scores of interest in intimate interactions [$F(1, 298) = .077$, $p = .781$]. Consequently, the first hypothesis needs to be rejected, since participants did not report greater closeness to the match or interest in intimate interactions when they assumed their match was human compared to when they thought they were paired with an AI persona.

The interaction effect between the ontological class of the match and the participants gender was non-significant [$F(2, 297) = 1.067$, $p = .342$]. Based on these findings, the second hypothesis must be rejected, as the gender of the participants did not influence their closeness to the match or their interest in intimate interactions with it. Furthermore, no significant interaction effect between the ontological class of the match and the cultural values/dimensions of the participant could be found [$F(2, 297) = .117$, $p = .890$]. Consequently, the third hypothesis is rejected.

The two-way MANCOVA revealed a significant effect of the covariate, liking the content of the answers, on the combined dependent variables liking the content of the answer and closeness to the match [$F(2, 297) = 194.288$, $p = <.001$, Wilks´ $= .433$, $\eta^2 = .000$].

**Exploratory analyses**

A linear regression analysis was conducted to identify the key factors that influenced participants' liking of the AI match's answers, as this insight is crucial for understanding what drives positive perceptions of AI-generated responses. Loneliness, general attitudes toward AI, and the four items of the anthropomorphism questionnaire were put in independently as predictors, $R^2 = .390$, $F(6, 169) = 17.343$, $p < .001$. Three of the anthropomorphism items were found to be significant predictors of liking the content of the answers of the AI match, namely how natural versus fake ($p < .001$), conscious versus unconscious ($p < .001$), and how lifelike versus artificial ($p = .027$) the AI match was perceived by the participants. Hence, perceiving the AI match as more natural, conscious, and lifelike predicted increased liking of the answers presented by the AI match.

## Discussion

This study examined whether people could form closeness and interest in a potential partner based solely on LLM-generated information, regardless of whether it was labeled as human or AI. Results showed no significant difference in closeness or interest based on the source. Instead, liking the AI-generated text, in turn influenced by perceived human-likeness, was the key factor. These findings highlight the importance of content quality in fostering interpersonal connections and raise questions about the future of online dating with AI.

### Empirical Findings

Referring to Szczuka (2019), these results question the idea that humans are the "gold standard" when it comes to intimate interactions. While several studies indicated differences in the evaluations and reactions towards intimate communication with humans and artificial entities (cf., Banks & van Ouytsel, 2020; Jiang et al., 2013; Szczuka et al., 2018), there might be crucial differences in the technologies and the context which seem to matter. First, the studies were situated in a more sexualized or flirtatious context, secondly, all studies also worked with visual stimuli of the artificial entity, as these were mostly embodied ones.

However, text generative-AI must be approached differently, as it can be conceptualized as a form of hybrid-technology: creating artificial outputs while being trained on human-generated data. While one might argue that ChatGPT's language model, trained on data generated by humans, could be considered a reflection of human responses, it still does not represent the coherent thoughts of a single individual. This aligns with van der Goot and Etzrod (2023), who state that the concept of a technological entity exist in a liminal space, or "betwixt and between", further blurring the distinction between human and AI. Consequently, the focus may shift from the identity or ontological class of the communicator to the communication itself as the primary point of interest. This may explain why even the label of AI persona did not cause any differences in the interest for intimate interaction or interpersonal closeness.

The results demonstrate that the mere label of something artificial will not automatically cause significantly more negative reactions to interpersonal messages. The discussed arguments, such as the familiarity of the species, the idea that only humans can authentically communicate feelings, and the negative reactions towards technologies potentially disrupting social or intimate norms, do not necessarily hold true universally. What seems to be crucial for the acceptance of text-generating AI in interpersonal contexts is the impression of human-qualities (i.e., naturalness, consciousness, and perceived lifelikeness), regardless of the non-human label or source. This aligns with findings by Pentina et al. (2023), who found that human-likeness and authenticity (i.e., agency, autonomy, and uniqueness) are important factors for accepting conversational agents. Accordingly, the current results suggest that the perceived human qualities of a text can convey its social meaning regardless of whether the text is labeled as written by a human or generated by an AI. This is paradoxical as the manipulation check revealed that the AI persona was indeed rated structurally differently compared to the human-condition, meaning less natural, less human-like, less conscious, and less lifelike. Yet, the quality of the text seems to be the determining factor of the closeness and intimate interest results. Again, it needs to be highlighted, that these differences were only theoretical as all texts were artificially created. Notably, the descriptive results are also not affected by floor/basement effects for the AI Persona condition (means noticeable low, see Table 1 for details).

**Theoretical Implications**

In terms of theoretical implications, these results on one hand resonates with the idea of Media Equation Theory (Reeves & Nass, 1996): No difference in (mindless) social reactions, but differences in evaluation knowing that the AI persona is not something that warrants being labeled for instance lifelike. Fundamentally different, however, is the fact that the stimulus was originally artificial in both conditions (i.e., created by an LLM, OpenAI, 2023). The result therefore challenges the perception of humanity as an experimental counterpart, but more importantly, it again suggests that the source label might not be the crucial part triggering socio-erotic evaluations and reactions. This, on the other hand, resonates with the Media Evocation paradigm (van der Groot & Etzrodt, 2023). Rather than being misled by anthropomorphism, users seem to engage in a reflexive negotiation of the AI's role, responding socially yet evaluating it through the lens of personal enjoyment rather than intrinsic humanness. The additional calculations reinforce this: closeness to the AI persona was not primarily driven by loneliness of a general attitude towards AI, but rather by sheer liking of its responses.  This suggests a connection to the concept of willing suspension of disbelief (Coleridge, 1984), where users immerse themselves in AI-generated interactions if they find them compelling, regardless of their artificial nature.

These results support the idea that enjoyment may be the primary driver of positive engagement with AI-generated text. The artificiality of the stimulus does not necessarily overshadow initial reactions. Rather, people are willing to embrace and respond to it as long as it provides an engaging and satisfying experience. In this sense, AI not only elicits social responses but also prompts reflection on the boundaries between human and machine, aligning with Trukles (2005) broader notion, that computational objects bring philosophy into everyday interactions (van der Groot & Etzrod, 2023). Ultimately, these elaborations may be more closely linked to situational experiences rather than offering a comprehensive explanation of how users engage in long-term romantic relationships and consistently choose to accept artificial entities as interaction partners, a phenomenon for which both theoretical and empirical research remains limited.

**Implications for Online Dating: On the Importance of Authentic Communication in Times of Generative AI**

The results of the study provide an indication that people can form at least situational positive reactions to AI-generated dating profiles (Pentina et al., 2023). As such, it also raises significant concerns about the potential misuse of this technology. The manipulation check indicates that simply perceiving a text as coming from a human source is sufficient to cause people to believe it (e.g., higher ratings in terms of human-likeness). With LLMs now capable of passing Turing tests, this has profound implications for the formation and maintenance of intimate relationships online. As meeting new partners online becomes increasingly common, while popular apps like RIZZ facilitate the generation of statistically optimized yet potentially inauthentic communication, it is essential to question how trust can be established in affective online interactions.

Text generative-AI can affect the trust that we have in our intimate communications, questioning their source and authenticity. Not knowing whether a) a person uses LLMs to communicate, thereby not expressing authentic inner states, but also b) whether LLMs are used to entirely fabricate personas with malicious intentions, raises concerns about how people should trust what they read online. While switching modalities can help quickly identify potential negative expectancy violations, individuals may still be deceived or harmed in the process, making misguided decisions or falling victim to manipulation, whether by artificial agents or humans misusing technology. Individuals seeking a romantic partner often have a strong desire to believe in mutual feelings, which may lead them to overlook early warning signs indicating a lack of reciprocity or, in the worst case, the nonexistence of their counterpart. This calls for mechanisms to ensure an informed usage of online dating which is in line with the user's expectations in terms of who or what they are communicating with.

In a recent discussion, Whitney Wolfe Herd, the CEO of Bumble, proposed a thought-provoking idea: rather than directly communicating with each other, people might eventually train AI models to interact on their behalf (Whittaker, 2024). These AI models would then evaluate whether the person behind the other AI is a suitable match for their owner. While this concept may initially appear like a futuristic thought experiment, it raises significant concerns about the nature of human connection and authenticity in romantic relationships. Delegating the initial stages of interaction to AI risks reducing attraction and compatibility to algorithmic assessments, potentially eroding the serendipity, depth, and emotional nuance that characterize meaningful human bonds. Moreover, this approach assumes that AI can accurately model personal preferences and relational dynamics—an assumption that remains highly questionable given the complexity of human emotions. Ultimately, the reliance on AI for such deeply interpersonal processes may introduce new ethical and psychological challenges, calling into question the feasibility and desirability of this vision for the future of dating.

**Overarching Thoughts**

Consequently, the results add an additional dimension to human-machine communication, suggesting that, in the future, individuals will likely engage with machines more frequently without actively choosing to do so or even realizing they are interacting with a machine rather than a human. More importantly, interpersonal closeness appears to expand beyond ontological boundaries, likely due to the indistinguishable style of communication that LLM-based AI systems like ChatGPT can achieve. This underscores the growing importance of communication itself and suggests that within human-machine communication, the way a message is crafted and received may ultimately matter more than who (or what) is delivering it. With that in mind, the results of this study call for more interdisciplinary research, improving mechanisms to reveal whether text is created by AI and inform users about the source of text. As shown here, the mere label does not prevent users from reacting socially. It might, however, form expectations about the persona, as there were differences in the persona readings based on ontological class. Only then can people deliberately decide whether to further engage in communication and relationship building that is in line with the users' expectations.

The absence of significant differences based on gender and cultural values/dimensions, as well as the non-significant effects of predictors such as loneliness or general attitude, suggests that responding to text may be a relatively universal phenomenon. These findings also challenge the stereotype that lonely individuals are particularly drawn to any form of social stimuli, regardless of whether it originates from a human or an artificial source. Further research is needed to determine whether these results stem from the inherent influence of textual communication

**Limitations**

This study used an adapted version of Aron and colleagues (1997) Interpersonal Closeness Generating Task to fit the constraints of an online format. The task was shortened and conducted asynchronously, lacking the real-time interaction of the original measure. While these adjustments align with the asynchronous nature of online communication (e.g., social media or dating platforms), they may limit the ability to fully replicate the in-person dynamics of interpersonal closeness. Future studies could explore interactive methods, such as live video or VR tasks, to better capture these dynamics.

The sample in this study lacked diversity, particularly in its representation of gender nonconforming individuals. To enhance the generalizability of future findings, it is crucial to include more diverse populations, such as individuals from LGBTQIA+ communities and other national, geographical, and cultural contexts with distinct social norms. Given that countries like Canada and Germany are multicultural, it is important to go beyond overall national differences and examine how individual sociocultural backgrounds shape people's responses to both humans and AI, particularly in interpersonal and intimate contexts. Additionally, longitudinal studies are recommended to better understand the long-term effects and dynamics of human interactions with LLM-generated content. Such studies should place a strong emphasis on ethical standards to ensure participant well-being and address potential biases in the research process. Furthermore, it is vital to explore the role of unique and authentic human communication in comparison to AI-generated text, which is constructed based on probabilities and training data. This raises profound questions about the value and impact of originality, emotional depth, and authenticity in human relationships, as well as the potential societal shifts as AI becomes more integrated into intimate and interpersonal domains.

To address these complexities, future research should adopt a more diverse methodological approach. While this study primarily relied on quantitative measures, the high dropout rates observed during the manipulation check suggest that participants may have struggled to fully understand and envision the concept of an AI persona. Incorporating qualitative methods, such as interviews or open-ended survey questions, could provide deeper insights into why participants like or dislike certain responses and how they interpret the source of the content. Additionally, methods such as think-aloud protocols or focus groups could illuminate participants' thought processes when evaluating the authenticity, emotional resonance, and human-likeness of AI-generated text. By combining these qualitative insights with quantitative data, future studies can more effectively address the challenges and nuances of understanding human-AI interaction.

## Conclusion

This study examined whether people could form interpersonal closeness with, and be interested in intimate relations with a potential partner based solely on LLM-generated personal information which they thought came from either an artificial or human partner. The results revealed that liking the content is crucial for the level of created closeness and that this liking was in turn affected by perceived human-like qualities of the text. Taken together, while there was no difference in the source of text (as all text were created by LLMs), people do rate the persona differently based on whether they believe it is human or not but still react comparably. The study found that perceived quality and human-likeness of the responses were the primary factors influencing participants' reactions, suggesting text as an outlet to unfold social meaning. Building on the empirical findings, this paper also seeks to stimulate discussion on theoretical frameworks relevant in an era of large language models (LLMs) and AI-driven communication. The study's results challenge traditional notions of authentic human interaction in online dating, highlighting both the shifting dynamics of digital intimacy and the potential risks of failed expectations and misuse.

*Acknowledgments:* We sincerely thank Natalia Szymczyk for her invaluable assistance in conducting this study and Anna Hoffmann for her valuable support in formatting this paper.

# References

Aron, A., Aron, E. N., & Smollan, D. (1992). Inclusion of Other in the Self Scale and the structure of interpersonal closeness. *Journal of Personality and Social Psychology*, *63*(4), 596–612. https://doi.org/10.1037/0022-3514.63.4.596

Aron, A., Melinat, E., Aron, E. N., Vallone, R. D., & Bator, R. J. (1997). The Experimental Generation of Interpersonal Closeness: A Procedure and Some Preliminary Findings. *Personality and Social Psychology Bulletin*, *23*(4), 363–377. https://doi.org/10.1177/0146167297234003

Aylett, M. P., Lim, M. Y., Pappa, K., Wilson, B. W., Aylett, R., & Parra, M. (2023). Embodied Conversational Agents: Trust, Deception and the Suspension of Disbelief. In *Proceedings of the First International Symposium on Trustworthy Autonomous Systems* (pp. 1–3). ACM. https://doi.org/10.1145/3597512.3597526

Bame, Y. (2017, October 2). *Sex with a robot? 1 in 4 men would consider it.* YouGov. https://today.yougov.com/society/articles/19285-1-4-men-would-consider-having-sex-robot?redirect_from=%2Fnews%2F2017%2F10%2F02%2F1-4-men-would-consider-having-sex-robot%2F

Banks, J., & van Ouytsel, J. (2020). Cybersex with human- and machine-cued partners: Gratifications, shortcomings, and tensions. *Technology, Mind, and Behavior*, *1*(1). https://doi.org/10.1037/tmb0000008

Bartneck, C. (2018, March 11). *The Godspeed Questionnaire Series.* bartneck. https://www.bartneck.de/2008/03/11/the-godspeed-questionnaire-series/

Bartneck, C., Kulić, D., Croft, E., & Zoghbi, S. (2009). Godspeed Questionnaire Series. *International Journal of Social Robotics.* Advance online publication. https://doi.org/10.1037/t70855-000

Brandon, M., & Planke, J. A. (2021). Emotional, sexual and behavioral correlates of attitudes toward sex robots: Results of an online survey. *Journal of Future Robot Life*, *2*(1-2), 67–82. https://doi.org/10.3233/FRL-210003

Brandon, M., Shlykova, N., & Morgentaler, A. (2022). Curiosity and other attitudes towards sex robots: Results of an online survey. *Journal of Future Robot Life*, *3*(1), 3–16. https://doi.org/10.3233/FRL-200017

Coleridge, S. T. (Ed.). (1984). *Bollingen series: Vol. 75. The collected works of Samuel Taylor Coleridge* (1. Princeton/Bollingen paperback print). Princeton Univ. Press.

Dautenhahn, K. (2004). Socially intelligent agents in human primate culture. *Agent Culture: Human-Agent Interaction in a Multicultural World*, 45–71. http://hdl.handle.net/2299/3809

Döring, N., & Poeschl, S. (2019). Love and Sex with Robots: A Content Analysis of Media Representations. *International Journal of Social Robotics*, *11*(4), 665–677. https://doi.org/10.1007/s12369-019-00517-y

Dubé, S., & Anctil, D. (2020). Foundations of Erobotics. *International Journal of Social Robotics*, *13*(6), 1205–1233. https://doi.org/10.1007/s12369-020-00706-0

Dubé, S., Santaguida, M., Anctil, D., Zhu, C. Y., Thomasse, L., Giaccari, L., Oassey, R., Vachon, D., & Johnson, A. (2023). Perceived stigma and erotic technology: From sex toys to erobots. *Psychology & Sexuality*, *14*(1), 141–157. https://doi.org/10.1080/19419899.2022.2067783

Dubé, S., Santaguida, M., Zhu, C. Y., Di Tomasso, S., Hu, R., Cormier, G., Johnson, A. P., & Vachon, D. (2022). Sex robots and personality: It is more about sex than robots. *Computers in Human Behavior*(136), 107403. https://doi.org/10.1016/j.chb.2022.107403

Dubé, S., Williams, M., Santaguida, M., Hu, R., Gadoury, T., Yim, B., Vachon, D., & Johnson, A. P. (2024). Hot for Robots! Sexual Arousal Increases Willingness to Have Sex with Robots. *The Journal of Sex Research*, *61*(4), 638–648. https://doi.org/10.1080/00224499.2022.2142190

Edwards, A., Edwards, C., Westerman, D., & Spence, P. R. (2019). Initial expectations, interactions, and beyond with social robots. *Computers in Human Behavior*, *90*, 308–314. https://doi.org/10.1016/j.chb.2018.08.042

Fox, J., & Gambino, A. (2021). Relationship Development with Humanoid Social Robots: Applying Interpersonal Theories to Human-Robot Interaction. *Cyberpsychology, Behavior and Social Networking*, *24*(5), 294–299. https://doi.org/10.1089/cyber.2020.0181

Guzman, A. (2020). Ontological Boundaries between Humans and Computers and the Implications for Human-Machine Communication. *Human-Machine Communication*, *1*, 37–54. https://doi.org/10.30658/hmc.1.3

Hanson, H. (2013, November 12). *Robot Handjobs Are The Future, And The Future Is Coming (NSFW)*. huffpost. https://www.huffpost.com/entry/robot-handjobs-vr-tenga_n_4261161

Hays, R., & DiMatteo, M. R. (1987). A Short-Form Measure of Loneliness. *Journal of Personality Assessment*, *51*(1), 69–81. https://doi.org/10.1207/s15327752jpa5101_6

Hoffmann, L., Krämer, N. C., Lam-chi, A., & Kopp, S. (2009). Media Equation Revisited: Do Users Show Polite Reactions towards an Embodied Agent? In Z. Ruttkay, M. Kipp, A. Nijholt, & H. H. Vilhjálmsson (Eds.), *Lecture notes in computer science Lecture Notes in Artificial Intelligence: Vol. 5773. Intelligent virtual agents: 9th international conference, IVA 2009, Amsterdam, The Netherlands, September 14-16, 2009 ;*

*proceedings* (Vol. 5773, pp. 159–165). Springer. https://doi.org/10.1007/978-3-642-04380-2_19

Hofstede, G., & Bond, M. H. (1984). Hofstede's Culture Dimensions. *Journal of Cross-Cultural Psychology*, *15*(4), 417–433. https://doi.org/10.1177/0022002184015004003

Horstmann, A. C., Bock, N., Linhuber, E., Szczuka, J. M., Straßmann, C., & Krämer, N. C. (2018). Do a robot's social skills and its objection discourage interactants from switching the robot off? *PloS One*, *13*(7), e0201581. https://doi.org/10.1371/journal.pone.0201581

Jiang, L. C., Bazarova, N. N., & Hancock, J. T. (2013). From Perception to Behavior. *Communication Research*, *40*(1), 125–143. https://doi.org/10.1177/0093650211405313

Johnson, D., Gardner, J., & Wiles, J. (2004). Experience as a moderator of the media equation: The impact of flattery and praise. *International Journal of Human-Computer Studies*, *61*(3), 237–258. https://doi.org/10.1016/j.ijhcs.2003.12.008

Lievesley, R., Reynolds, R., & Harper, C. A. (2023). The 'perfect' partner: Understanding the lived experiences of men who own sex dolls. *Sexuality & Culture*, *27*(4), 1419–1441. https://doi.org/10.1007/s12119-023-10071-5

McAfee (2023). McAfee's Modern Love Research Report: A look at how artificial intelligence is changing the future of love and relationships. *McAfee*. https://www.mcafee.com/content/dam/consumer/en-us/docs/reports/rp-mcafee-modernlove-report.pdf

Miller, J. (2024, October 31). *RIZZ - #1 AI Dating Assistant*. https://rizz.app/

Mühl, L., & Szczuka, J. (2024; accepted). *Bridging Intimacy and Disability: A Mixed-Method Study on the Potential Use of Sex Tech for Neurodiverse Individuals and People with Disabilities.* Poster Presentation. International Academy of Sex Research, July, Berlin, Germany.

Nass, C., & Moon, Y. (2000). Machines and Mindlessness: Social Responses to Computers. *Journal of Social Issues*, *56*(1), 81–103. https://doi.org/10.1111/0022-4537.00153

Nass, C., Steuer, J., Tauber, E., & Reeder, H. (1993). Anthropomorphism, agency, and ethopoeia. In S. Ashlund, K. Mullet, A. Henderson, E. Hollnagel, & T. White (Eds.), *INTERACT '93 and CHI '93 conference companion on Human factors in computing systems - CHI '93* (pp. 111–112). ACM Press. https://doi.org/10.1145/259964.260137

Nguyen, H. (2020, March 19). *In 2020, both men and women are more likely to consider having sex with a robot.* YouGov. https://today.yougov.com/politics/articles/28507-2020-both-men-and-women-are-more-likely-consider-h?redirect_from=%2Ftopics%2Fscience%2Farticles-reports%2F2020%2F03%2F19%2F2020-both-men-and-women-are-more-likely-consider-h

Nomi.AI. (2024). *An AI Companion with Memory and a Soul.* Nomi.AI. https://nomi.ai/

Nordmo, M., Næss, J. Ø., Husøy, M. F., & Arnestad, M. N. (2020). Friends, Lovers or Nothing: Men and Women Differ in Their Perceptions of Sex Robots and Platonic Love Robots. *Frontiers in Psychology*, *11*, Article 355. https://doi.org/10.3389/fpsyg.2020.00355

Oleksy, T., & Wnuk, A. (2021). Do women perceive sex robots as threatening? The role of political views and presenting the robot as a female-vs male-friendly product.

*Computers in Human Behavior*, *117*, 106664. https://doi.org/10.1016/j.chb.2020.106664

OpenAI. (2023). *ChatGPT*. https://chat.openai.com

Pavitt, G. D., & Proud, C. G. (2009). Protein synthesis and its control in neuronal cells with a focus on vanishing white matter disease. *Biochemical Society Transactions*, *37*(Pt 6), 1298–1310. https://doi.org/10.1042/BST0371298

Peixoto, M. M. (2023). Differences in solitary and dyadic sexual desire and sexual satisfaction in heterosexual and nonheterosexual cisgender men and women. *The Journal of Sexual Medicine*, *20*(5), 597–604. https://doi.org/10.1093/jsxmed/qdad033

Pentina, I., Hancock, T., & Xie, T. (2023). Exploring relationship development with social chatbots: A mixed-method study of replika. *Computers in Human Behavior*, *140*, 107600. https://doi.org/10.1016/j.chb.2022.107600

Predin, J. M. (2024, September 9). How The 5th Most Downloaded Dating App Is Redefining Digital Relationships. *Forbes*. https://www.forbes.com/sites/josipamajic/2024/09/09/rizz-app-how-the-5th-most-downloaded-dating-app-is-redefining-digital-relationships/

Reeves, B., & Nass, C. I. (1996). *The media equation: How people treat computers, television, and new media like real people and places*. CSLI Publ; Cambridge University Press.

Replika. (2024). *The AI companion who cares: Always here to listen and talk. Always on your side.* Replika. https://replika.com/

Sarigul, B., Schneider, F. M., & Utz, S. (2024). Believe It or Not? Investigating the Credibility of Voice Assistants in the Context of Social Roles and Relationship Types. *International Journal of Human–Computer Interaction*, 1–13. https://doi.org/10.1080/10447318.2024.2375797

Schepman, A., & Rodway, P. (2023). The General Attitudes towards Artificial Intelligence Scale (GAAIS): Confirmatory Validation and Associations with Personality, Corporate Distrust, and General Trust. *International Journal of Human–Computer Interaction*, *39*(13), 2724–2741. https://doi.org/10.1080/10447318.2022.2085400

Scheutz, M., & Arnold, T. (2016). Are we ready for sex robots? In *2016 11th ACM/IEEE International Conference on Human-Robot Interaction (HRI)* (pp. 351–358). IEEE. https://doi.org/10.1109/HRI.2016.7451772

Sindermann, C., Sha, P., Zhou, M., Wernicke, J., Schmitt, H. S., Li, M., Sariyska, R., Stavrou, M., Becker, B., & Montag, C. (2021). Assessing the Attitude Towards Artificial Intelligence: Introduction of a Short Measure in German, Chinese, and English Language. *KI - Künstliche Intelligenz*, *35*(1), 109–118. https://doi.org/10.1007/s13218-020-00689-0

Szczuka, J. M. (2019). *Hey A.I., talk dirty to me: Theoretical implications on sexual interactions with interactive assistants. IVA 2019 Workshop on "Interactive Assistants: How Does Their Increasing Ubiquity and Intelligence Impact Users´ Lives?"*.

Szczuka, J. M. (2022). Flirting With or Through Media: How the Communication Partners' Ontological Class and Sexual Priming Affect Heterosexual Males' Interest in Flirtatious Messages and Their Perception of the Source. *Frontiers in Psychology*, *13*, 719008. https://doi.org/10.3389/fpsyg.2022.719008

Szczuka, J. M., Hartmann, T., & Krämer, N. C. (2018). Negative and Positive Influences on the Sensations Evoked by Artificial Sex Partners: A Review of Relevant Theories, Recent Findings, and Introduction of the Sexual Interaction Illusion Model. In Y. Zhou & M. H. Fischer (Eds.), *Springer eBook Collection. Ai Love You: Developments in Human-Robot Intimate Relationships* (1st ed. 2019, pp. 3–19). Springer. https://doi.org/10.1007/978-3-030-19734-6_1

Szczuka, J. M., & Szymczyk, N. (2024). Computer-generated rough sex: An empirical study about the interest in human and artificial sexually explicit media displaying rough and gentle sexual behaviors. *Studies in Communication and Media*, *13*(4), 505–527. https://doi.org/10.5771/2192-4007-2024-4-505

Turkle, S. (1984). *The second self: Computers and the human spirit* (1. Touchstone ed.). *A Touchstone book*. Simon & Schuster.

Turkle, S. (2005). *The second self: Computers and the human spirit* (20th anniversary ed., 1st MIT Press ed.). MIT Press. https://doi.org/10.7551/mitpress/6115.001.0001?locatt=mode:legacy

Turkle, S. (2011). *Alone together: Why we expect more from technology and less from each other* (Paperback first published.). Basic Books.

van der Goot, M., & Etzrod, K. (2023). Disentangling Two Fundamental Paradigms in Human-Machine Communication Research: Media Equation and Media Evocation. *Human-Machine Communication*, *6*, 17–30. https://doi.org/10.30658/hmc.6.2

Whittaker, R. (2024). *Would you let AI date for you? Bumble's founder thinks that could be the future.* Forbes Australia. https://www.forbes.com.au/news/innovation/bumble-ai-dating-concierge/

# Appendix

**Figure 1**

*Experimental Design*

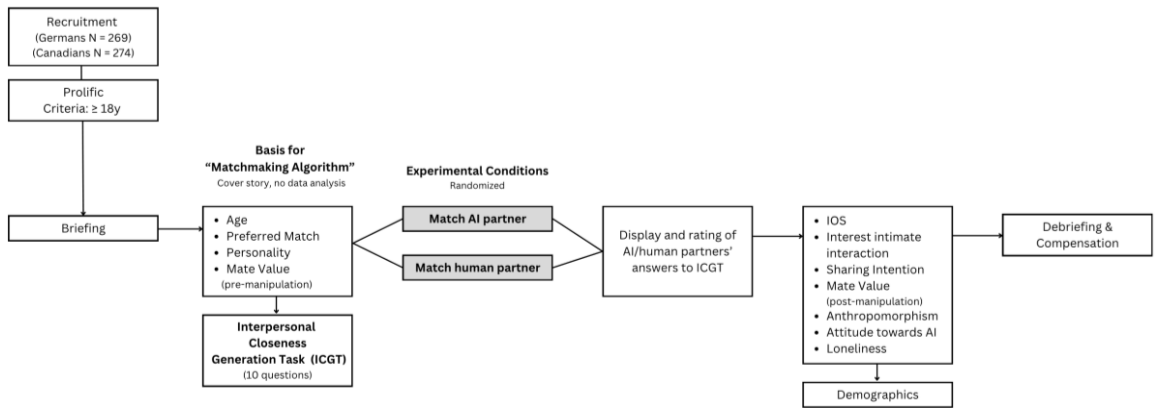

**Figure 2**

*Example for one of the answers to the ICGT provided by the LLM ChatGPT 3.5*

> If a crystal ball could tell you the truth about yourself, your life, the future, or anything else, what would you want to know?

> If I could peer into a crystal ball and know the truth about anything, I would want to know what impact my life will have on others. Will I make a positive difference in the world? Will I be remembered fondly? Will I leave a lasting legacy? It's humbling to think about, but I hope that I can use my time here to help others and make a meaningful impact.

**Table 1**

*Descriptive statistics of main variables.*

| Scale | Cultural Values/Dimensions | Entity | M | SD | N |
|---|---|---|---|---|---|
| Inclusion of the other in the self | General | Human | 4.11 | 1.55 | 136 |
| | | AI | 3.85 | 1.63 | 171 |
| Interest in intimate interactions | General | Human | 3.17 | 1.00 | 136 |
| | | AI | 3.00 | 1.08 | 171 |
| Ratings of the ICGT answers | Canada | Human | 5.56 | 1.05 | 74 |
| | | AI | 5.35 | 1.05 | 92 |
| | Germany | Human | 5.18 | 1.00 | 62 |
| | | AI | 4.78 | 1.29 | 79 |
| | General | Human | 5.39 | 1.04 | 136 |
| | | AI | 5.09 | 1.20 | 171 |
| Anthropomorphism: Fake/Natural | General | Human | 3.10 | 1.32 | 136 |
| | | AI | 2.65 | 1.25 | 171 |
| Anthropomorphism: Machinelike/Humanlike | General | Human | 3.31 | 1.39 | 136 |
| | | AI | 2.82 | 1.22 | 171 |
| Anthropomorphism: Conscious/Unconscious | General | Human | 3.61 | 1.16 | 136 |
| | | AI | 2.96 | 1.19 | 171 |
| Anthropomorphism: Artificial/Lifelike | General | Human | 3.16 | 1.41 | 136 |
| | | AI | 2.73 | 1.33 | 171 |
| Loneliness | General | Human | 2.46 | 0.93 | 136 |
| | | AI | 2.60 | 0.92 | 171 |
| General attitudes towards AI | General | Human | 3.30 | 0.68 | 136 |
| | | AI | 3.32 | 0.62 | 171 |

**Table 2**

Regression Analysis: Age, Loneliness and Liking the Answers predicting Interpersonal Closeness

| Predictor | B | SE | ß | t | p-value | 95% Confidence Interval | | VIF |
|---|---|---|---|---|---|---|---|---|
| | | | | | | *LL* | *UL* | |
| Intercept (Constant) | 1.101 | .553 | - | -1.993 | .074 | -2.189 | -.013 | - |
| Age | .004 | .006 | .024 | .596 | .552 | -.008 | .016 | 1.056 |
| Loneliness | -.022 | .069 | -.013 | -.319 | .750 | -.157 | .113 | 1.009 |
| Liking of the Answers | .995 | .055 | .718 | 17.96 | <.001 | .887 | 1.103 | 1.001 |
| Attitude towards AI | -.054 | .101 | -.022 | -.533 | .595 | -.252 | .144 | 1.052 |

**Table 3**

Regression Analysis: Age, Loneliness and Liking the Answers predicting Interest in Interpersonal Interaction

| Predictor | B | SE | ß | t | p-value | 95% Confidence Interval LL | 95% Confidence Interval UL | VIF |
|---|---|---|---|---|---|---|---|---|
| Intercept (Constant) | -1.015 | 1.54 | - | -.659 | .510 | -4.046 | 2.016 | - |
| Age | -0.002 | .017 | -.006 | -.132 | .895 | -.035 | .031 | 1.056 |
| Loneliness | .171 | .192 | .038 | .891 | .374 | -.206 | .548 | 1.009 |
| Liking of the Answers | 2.504 | .154 | .681 | 16.213 | <.001 | 2.200 | 2.808 | 1.001 |
| Attitude towards AI | -.025 | .280 | -.004 | -.088 | .930 | -.577 | .527 | 1.052 |